\documentclass[12pt]{article}
\usepackage{amssymb}

\begin{document}

\bigskip

\bigskip\ 

\begin{center}
\textbf{SEARCH FOR A ``GRAVITOID'' THEORY}

\bigskip\ 

\smallskip\ 

J. A. Nieto\footnote[1]{%
nieto@uas.uasnet.mx} and M. C. Mar\'{i}n

\smallskip\ 

\textit{Departamento de Investigaci\'{o}n en F\'{i}sica de la Universidad de
Sonora,}

\textit{83190, Hermosillo Sonora, M\'{e}xico}

\textit{and}

\textit{Facultad de Ciencias F\'{i}sico-Matem\'{a}ticas de la Universidad
Aut\'{o}noma}

\textit{de Sinaloa, 80010, Culiac\'{a}n Sinaloa, M\'{e}xico}\footnote[2]{%
permanent address}

\bigskip\ 

\bigskip\ 

\textbf{Abstract}
\end{center}

By combining the concepts of graviton and matroid, we outline a new
gravitational theory which we call gravitoid theory. The idea of this theory
emerged as an attempt to link the mathematical structure of matroid theory
with M-theory. Our observations are essentially based on the formulation of
matroid bundle due to MacPherson and Anderson-Davis. Also, by considering
the oriented matroid theory, we add new observations about the link between
the Fano matroid and $D=11$ supergravity which was discussed in some of our
recent papers. In particular we find a connection between the affine matroid 
$AG(3,2)$ and the $G_{2}-$symmetry of $D=11$ supergravity.

.\bigskip

\bigskip

\bigskip\ 

Pacs numbers: 04.60.-m, 04.65.+e, 11.15.-q, 11.30.Ly

February, 2003

\newpage\ 

\noindent \textbf{I. INTRODUCTION}

\bigskip

It is known that the theory called M-theory$^{1-3}$ was suggested by duality
symmetries interrelating the five known superstring theories$^{4}$ in 9+1
dimensions (i.e., nine space and one time), Type I, Type IIA, Type IIB,
Heterotic SO(32) and Heterotic E$_{8}\times $ E$_{8}.$ If the strong/weak
coupling duality transformations allow to relate these superstrings theories
then it is no longer appropriate to see them as distinct theories, but
rather as different manifestations of an underlying unique theory: M-theory.
Roughly speaking, we can say that duality implies M-theory.

It turns out that in an apparent complete different area and in another time
an analogue conclusion led Whitney$^{5}$ to the discovery of the concept of
matroid in 1935$.$ In effect, it is known that not all of the graphs have an
associated dual graph. If one insists on extending the duality property to
any graph, not only for plane graphs, one is inevitably led to the
mathematical structure of matroids. From this point of view, we can also say
that duality implies a matroid scenario.

Now, the arising question is whether these two conclusions are not a
coincidence. But, why to ask, first of all, for a coincidence if M-theory
and matroid structure appear, at first sight, as two totally different
formalisms. This scenario seems to be similar to the case of the two
apparently unrelated subjects: Newton's gravitational theory and the
mathematics of Riemann geometry. The equivalence principle, among other
things, suggested to look for a mathematical structure beyond the Euclidean
geometry used in Newton's gravitational theory. It is well known that in
1915 it was discovered that the use of the Riemann geometry makes possible
the accomplishment of such a quest. Similarly, a duality principle in
M-theory may suggest the search of a singular mathematical structure beyond
the mathematical formalism used in string theory. In three previous works$%
^{6-8}$ we provided enough evidence to prove that a duality principle in
M-theory may have a mathematical support in the well-developed formalism of
matroid-theory$^{9-13}$. In fact, in these references it was shown that Fano
matroid and the matroid-Tutte polynomial are connected to $D=11$
Supergravity, Chern-Simons theory and string theory, which are essential
parts of M-Theory.

Further, the connection between matroid theory and $D=11$ supergravity may
naturally lead to the need of a mathematical structure that combines
matroids and fiber bundles. The reason for this is that in order to have a
realistic four dimensional theory it is necessary to compactify seven
dimensions out of the eleven space-like dimensions of $D=11$ supergravity.
The traditional method to achieve such a compactification is via the
Kaluza-Klein mechanism. But, as it is well known, this mechanism is closely
related to the mathematical structure of fiber bundles.

Fortunately, the concept of matroid bundle has already been developed by the
mathematicians. In fact, in 1993 MacPherson$^{14}$ defined for the first
time matroid bundles in connection with combinatorial differential
manifolds. Since then the matroid bundles structure has been refined and
developed by Anderson and Davis,$^{15,25}$ and others$^{16-17}.$ Roughly
speaking, matroid bundles are the combinatorial analogue of vector bundles.
The central idea in matroid bundles is to allow oriented matroids$^{18}$ to
play the role of tangent spaces or, more generally, vector spaces.

Now, vector bundles provide an important tool in the formulation of the four
dimensional gravitational theory contained in $D=11$ supergravity.
Therefore, one should expect that matroid bundles may lead to a new
gravitational structure. We call this expected new gravitational object the
``gravitoid'' because it combines the concepts of graviton and matroid. Our
conjecture is that the gravitoid theory should play a crucial role not only
in $D=11$ supergravity but also in M-theory itself. Since every matroid has
an associated unique dual it is reasonable to expect that the gravitoid
should have an associated unique dual gravitoid and this fact should lead to
a dual invariant partition function for the gravitoid.

The plan of this work is as follows: In section II, we briefly review
oriented matroid theory. In section III, we closely follow the Ref. 6 adding
new information about the connection between matroid theory and $D=11$
supergravity. In particular, we discuss a link between the affine matroid $%
AG(3,2)$ and $G_{2}-$symmetry of $D=11$ supergravity via self-duality. In
section IV, we introduce the concept of matroid bundle and develop the
gravitoid theory. Finally, in section V, we remark the importance of
gravitoid theory and matroid bundles for further research and make some
final comments.

\bigskip

\smallskip\ 

\noindent \textbf{II. A BRIEF REVIEW OF ORIENTED MATROIDS}

\bigskip

Matroid theory$^{9-13}$ arose in 1935 from the abstract properties of vector
spaces and graph theory. Whitney$^{5}$ seems to be the first mathematician
to have noticed such properties. Looking at Whitney's previous work$^{19}$
on non-separable graph theory, one wonders if the lack of duality on
nonplanar graphs, such as the complete graph $K_{5}$ and the bipartite graph 
$K_{3,3},$ provided his main motivation to introduce such abstract
properties. Indeed, as it is known, in graph theory only planar graphs have
an associated dual graph. As the Kuratowski's theorem assures, the complete
graph $K_{5}$ and the bipartite graph $K_{3,3},$ which are nonplanar, do not
have a corresponding dual graph. On the other hand, matroid theory has the
attractive feature that every matroid has an associated unique dual matroid.
In particular, the matroid associated to $K_{5}$, let us say $M(K_{5})$, has
the dual matroid $M^{\ast }(K_{5})$, but $M^{\ast }(K_{5})$ turns out to be
not representable by means of a graph.

In connection with vector spaces, the matroid duality is realized by the
orthogonality of matrices. But the notion of orthogonality in vector spaces
is extended by matroid theory in the sense that not all matroids are
realizable by matrices. In fact, there are matroids, such as the non-Pappus
matroid,$^{9}$ that are not realizable by a subset of a vector space or by
matrices.

A natural generalization of graph theory is provided by oriented graph
theory. In this generalized case, the edges of a graph are labeled with a
plus or minus sign. Similarly, a natural generalization of matroid theory is
the oriented matroid theory. The origin of oriented matroid theory can be
traced back to the Gutierrez-Novoa's work$^{20}$on signed bases, in 1965.
However, it seems that the formal axiomatic notion of orientability for
matroids was not proposed before 1978 (see Ref. 18 for details). One of the
attractive features of oriented matroids is that they capture the sign
properties of vector spaces without losing the duality properties of the
underlying non-oriented matroids. At present, oriented matroid theory is a
well-developed mathematical structure in combinatorics with a very
interesting connection in differential topology, as we shall discuss in
section IV.

An oriented matroid $\mathcal{M}$ is a pair $(E,\mathcal{C})$, where $E$ is
a non-empty finite set, and $\mathcal{C}$ is a non-empty collection of
subsets of $E$ (called signed circuits) satisfying the following properties:

$(\mathcal{C}\mathit{i})$ no circuit properly contains another circuit,

$(\mathcal{C}\mathit{ii)}$\textit{\ }if $\mathcal{C}_{1}$ and $\mathcal{C}%
_{2}$ are two distinct signed circuits, $\mathcal{C}_{1}\neq -\mathcal{C}%
_{2},$ and $c\in \mathcal{C}_{1}^{+}\cap \mathcal{C}_{2}^{-}$ then there
exists a third circuit $\mathcal{C}_{3}\in \mathcal{C}$ with $\mathcal{C}%
_{3}^{+}\subseteq (\mathcal{C}_{1}^{+}\cap \mathcal{C}_{2}^{+})\backslash
\{c\}$ and $\mathcal{C}_{3}^{-}\subseteq (\mathcal{C}_{1}^{-}\cap \mathcal{C}%
_{2}^{-})\backslash \{c\}$.

An alternative, but equivalent, way to define an oriented matroid is as
follows: An oriented matroid $\mathcal{M}$ is a pair $(E,\chi ),$ where $E$
is a non-empty finite set and $\chi $ (called chirotope) is a mapping $%
E^{r}\rightarrow \{-1,0,1\}$, with $r$ the rank on $E$, satisfying the
following properties.

$(\chi i)$ $\chi $ is not identically zero,

$(\chi ii)\chi $ is alternating,

$(\chi iii)$ for all $x_{1},x_{2},...,x_{r}$ and $y_{1},y_{2},...,y_{r}$
such that

\begin{equation}
\chi (x_{1},x_{2},...,x_{r})\chi (y_{1},y_{2},...,y_{r})\neq 0,  \label{1}
\end{equation}
there exists an $i\in \{1,2,...,r\}$ such that

\begin{equation}
\chi (y_{i},x_{2},...,x_{r})\chi
(y_{1},y_{2},...,y_{i-1},x_{1},y_{i+1,}...,y_{r})=\chi
(x_{1},x_{2},...,x_{r})\chi (y_{1},y_{2},...,y_{r}).  \label{2}
\end{equation}

For a vector configuration the chirotope $\chi $ can be identified as

\begin{equation}
\chi (i_{1},...,i_{r})\equiv sign\det (v_{i_{1}},...,v_{i_{r}})\in \{-1,0,1\}
\label{3}
\end{equation}
and in this case (2) becomes connected with the Grassmann-Plucker relation.

It can be proved that the definition of the underlying matroid $M$ of $%
\mathcal{M}$ follows from the chirotope definition for oriented matroids. In
fact, from the chirotope definition it follows that if $\mathcal{B}$ is the
set of $r$-subsets of $E$ such that

\begin{equation}
\chi (x_{1},x_{2},...,x_{r})\neq 0,  \label{4}
\end{equation}
for some ordering of $(x_{1},x_{2},...,x_{r})$ of $\mathcal{B}$, then $%
\mathcal{B}$ is the set of bases of the matroid $M$. Formally, the
definition of $M$ in terms of the bases is as follows:

A matroid $M$ is a pair $(E,\mathcal{B})$, where $E$ is a non-empty finite
set and $\mathcal{B}$ is a non-empty collection of subsets of $E$ (called
bases) satisfying the following properties:

$(\mathcal{B}$ $\mathit{i)}$\textit{\ }no base properly contains another
base;

$(\mathcal{B}$ $\mathit{ii)}$ if $B_{1}$ and $B_{2}$ are bases and if $b$ is
any element of $B_{1},$ then there is an element $g$ of $B_{2}$ with the
property that $(B_{1}-\{b\})\cup \{g\}$ is also a base.

The dual of $M$, denoted by $M^{\ast },$ is defined as a pair $(E,\mathcal{B}%
^{\ast }),$ where $\mathcal{B}^{\ast }$ is a non-empty collection of subsets
of $E$ formed with the complements of the bases of $M$. An immediate
consequence of this definition is that every matroid has a dual and this
dual is unique. It also follows that the double-dual $M^{\ast \ast }$ is
equal to $M$.

The definition of the corresponding dual oriented matroid $\mathcal{M}^{\ast
}$ is straightforward. One introduces the dual chirotope $\chi ^{\ast }$
such that

\begin{equation}
\chi ^{\ast }:E^{n-r}\rightarrow \{-1,0,1\}  \label{5}
\end{equation}
and

\begin{equation}
(x_{1},x_{2},...,x_{n-r})\rightarrow \chi (x_{1}^{\prime },x_{2}^{\prime
},...,x_{r}^{\prime })sign(x_{1},x_{2},...,x_{n-r},x_{1}^{\prime
},x_{2}^{\prime },...,x_{r}^{\prime }),  \label{6}
\end{equation}
where $(x_{1}^{\prime },x_{2}^{\prime },...,x_{r}^{\prime })$ means some
permutation of $E\backslash (x_{1},x_{2},...,x_{n-r})$ and

\[
sign(x_{1},x_{2},...,x_{n-r},x_{1}^{\prime },x_{2}^{\prime
},...,x_{r}^{\prime }) 
\]
is the parity of the number of inversions of $(1,2,...,n).$ It is not
difficult to see that, as in the case of ordinary matroids, every oriented
matroid $\mathcal{M}(E,\chi )$ has an associated unique dual $\mathcal{M}%
^{\ast }(E,\chi ^{\ast }).$ Furthermore, it is found that $\mathcal{M}^{\ast
\ast }=\mathcal{M}$.

Another interesting and important result is that

\begin{equation}
(\mathcal{M}_{1}\oplus \mathcal{M}_{2})^{\ast }=\mathcal{M}_{1}^{\ast
}\oplus \mathcal{M}_{2}^{\ast },  \label{7}
\end{equation}
where $\mathcal{M}_{1}\oplus \mathcal{M}_{2}$ is the direct sum of two
connected oriented matroids $\mathcal{M}_{1}$ and $\mathcal{M}_{2}$.

\bigskip

\smallskip\ 

\noindent \textbf{III. ORIENTED MATROIDS, OCTONIONS, }$G_{2}$\textbf{%
-SYMMETRY AND }$D=11$ \textbf{SUPERGRAVITY}

\bigskip

In this section, we shall extend the discussion presented in Ref.6, 7 and 8
on the connection between matroid theory and $D=11$ supergravity. In
particular, we show that the complementary relation in seven dimensions
between the Fano matroid $F_{7}$ and octonions can be extended to eight
dimensions by means of the affine matroid $AG(3,2)$. We also prove that the
self-duality condition on the octonionic structure in eight dimensions
(necessary to account for the $G_{2}$-symmetry in $D=11$ supergravity) is in
concordance with the self-duality of $AG(3,2)$. At the same time, the idea
of this section is to prepare the motivation and notation to consider
matroid bundles and the gravitoid theory in the next section.

It is known that in order to get a realistic four dimensional theory from $%
D=11$ supergravity it is necessary to compactify seven space-like dimensions
out of the eleven dimensions. At this respect, the Englert's compactification%
$^{21}$ of $D=11$ supergravity provides us with one of the most attractive
solutions. In this case, the $D=11$ manifold $B^{11}$ is compactified in the
form $B^{4}\times S^{7}$, where $B^{4}$ is a four dimensional space-time
manifold and $S^{7}$ is the seven dimensional sphere. Such a solution turns
out to be connected to the parallelizable properties of $S^{7}$, which in
turn are related to octonions.$^{22}$ Moreover, in the Englert%
\'{}%
s mechanism of compactification the mapping $M^{11}\rightarrow M^{4}\times
S^{7}$ is made via the Kaluza-Klein formalism which mathematically can be
understood in terms of fiber bundles.

The Fano matroid $F_{7}$ is one of the most important on matroid theory
essentially because, among other things, it is the only minimal binary
irregular matroid. The ground set for $F_{7}$ is

\begin{equation}
E=\{1,2,3,4,5,6,7\}.  \label{8}
\end{equation}
The bases of $F_{7}$ are all those subsets of $E$ with three elements except
the seven subsets $f_{1}=\{1,2,3\},$ $f_{2}=\{5,1,6\},$ $f_{3}=\{6,4,2\},$ $%
f_{4}=\{4,3,5\},$ $f_{5}=\{4,7,1\},$ $f_{6}=\{6,7,3\}$ and $f_{7}=\{5,7,2\}$%
. The dual $F_{7}^{\ast }$ of $F_{7}$ can be obtained from the complements
of the bases of $F_{7}$.

In order to link $F_{7}$ with octonions let us introduce the set

\begin{equation}
\mathcal{E}=\{e_{1},e_{2},e_{3,}e_{4,}e_{5},e_{6},e_{7}\}.  \label{9}
\end{equation}
We shall denote by $p_{1}=\{e_{1},e_{2},e_{3}\}$, $p_{2}=\{e_{5},e_{1},e_{6}%
\}$, $p_{3}=\{e_{6},e_{4},e_{2}\}$, $p_{4}=\{e_{4},e_{3},e_{5}\}$, $%
p_{5}=\{e_{4},e_{7},e_{1}\}$, $p_{6}=\{e_{6},e_{7},e_{3}\}$ and $%
p_{7}=\{e_{5},e_{7},e_{2}\}$ the corresponding subsets $f_{i}$, with $%
i=1,2,...7$. Let us write an octonion in the form

\begin{equation}
q=q_{0}e_{0}+q_{1}e_{1}+q_{2}e_{2}+q_{3}e_{3}+q_{4}e_{4}+q_{5}e_{5}+q_{6}e_{6} 
+q_{7}e_{7},
\label{10}
\end{equation}
where $e_{0}$ denotes the identity. The imaginary units $e_{i}$ determine
the product of octonions through the formula

\begin{equation}
e_{i}e_{j}=-\delta _{ij}+\psi _{ij}^{k}e_{k},  \label{11}
\end{equation}
where $\delta _{ij}$ is the Kronecker delta and $\psi _{ijk}=$ $\psi
_{ij}^{l}\delta _{lk}$ are the fully antisymmetric structure constants. By
taking the $\psi _{ijk}$ equal to $1$ or $-1$ for each of the seven
combinations $p_{i}$ we can obtain every value of $\psi _{ijk}$.

The basic associator of any three imaginary units $e_{i},e_{j}$ and $e_{k}$
is

\begin{equation}
\langle e_{i},e_{j},e_{k}\rangle
=(e_{i}e_{j})e_{k}-e_{i}(e_{j}e_{k})=\varphi _{ijkm}e_{m},  \label{12}
\end{equation}
where $\varphi _{ijkl}$ is a fully antisymmetric four index tensor. This
means that the octonion (Cayley) algebra is not associative and in fact it
is an alternative algebra. It turns out that $\varphi _{ijkl}$ and $\psi
_{ijk}$ are dual to each other. Specifically, they are related by

\begin{equation}
\varphi _{ijkl}=(1/3!)\epsilon _{ijklmnr}\psi _{mnr},  \label{13}
\end{equation}
where $\epsilon _{ijklmnr}$ is the fully antisymmetric Levi-Civita tensor,
with $\epsilon _{12...7}=1$.

Anticipating a connection between $F_{7}$ and octonions we observe that
giving the numerical values $f_{i}$ for the indices of $\psi _{mnr}$ and
using (13) we get the other seven subsets of $E$ with four elements used to
define $F_{7}^{\ast }.$ For instance, if we consider $f_{1}$ then we have $%
\psi _{123}$ and (13) gives $\varphi _{4567}$ which leads to the circuit
subset $\{4,5,6,7\}$ of $F_{7}^{\ast }.$

Now, consider the matrix

\begin{equation}
A=\left( 
\begin{array}{ccccccc}
1 & 0 & 1 & 0 & 1 & 0 & 1 \\ 
1 & 1 & 0 & 0 & 0 & 1 & 1 \\ 
0 & 1 & 1 & 1 & 0 & 0 & 1
\end{array}
\right) .  \label{14}
\end{equation}
If we denote by $v_{i}$ the columns of this matrix, it is found that the
bases of the matrix $A$ are those subsets with three columns except the
subsets $h_{1}=\{v_{1},v_{2},v_{3}\}$, $h_{2}=\{v_{5},v_{1},v_{6}\}$, $%
h_{3}=\{v_{6},v_{2},v_{4}\}$, $h_{4}=\{v_{4},v_{3},v_{5}\}$, $%
h_{5}=\{v_{4},v_{7},v_{1}\}$, $h_{6}=\{v_{6},v_{7},v_{3}\}$ and $%
h_{7}=\{v_{5},v_{7},v_{2}\}.$ This shows that $F_{7}$ is not only realizable
by the matrix $A$ but also that it is a binary matroid. Actually, we have
performed the map $\varphi (i)=v_{i}$ and consequently the ground set $%
\{1,2,3,4,5,6,7\}$ becomes $\{v_{1},v_{2},v_{3},v_{4},v_{5},v_{6},v_{7}\}$
and the subsets $f_{i}$ are mapped to $h_{i}$

In principle, we can associate the chirotope

\begin{equation}
\chi (i_{1},i_{2},i_{3})=sign\det (v_{i_{1}},v_{i_{2}},v_{i_{3}})  \label{15}
\end{equation}
to $F_{7}$. But one of the intriguing properties of $F_{7}$ is that it is
not orientable matroid. Without formally proving this fact (see Ref. 9 for a
formal proof), one can verify by hand that $\chi (i_{1},i_{2},i_{3})$ does
not satisfy property $(\chi iii)$. Anyhow, a relationship between $\chi
(i_{1},i_{2},i_{3})$ and the structure constants for octonions $\psi
_{i_{1}i_{2}i_{3}}$ can be accomplished by means of the formula

\begin{equation}
\psi _{i_{1}i_{2}i_{3}}+\chi (i_{1},i_{2},i_{3})=C_{i_{1}i_{2}i_{3}},
\label{16}
\end{equation}
where $C_{i_{1}i_{2}i_{3}}\in \{-1,1\}$ can be identified with the chirotope
of the uniform matroid $U_{3,7}$ which is an excluded minor for $GF(5)$%
-representability, where $GF(q)$ denotes a finite field of order $q.$ The
expression (16) shows that $\psi _{i_{1}i_{2}i_{3}}$ and $\chi
(i_{1},i_{2},i_{3})$ determine $C_{i_{1}i_{2}i_{3}}$. This means that $F_{7}$
and the octonions are complementary concepts of the oriented uniform matroid 
$\mathcal{M}(U_{3,7})$ structure.

In order to understand further the formula (16) let us compute $\chi
(i_{1},i_{2},i_{3})$ for the different values of the indices $i_{p}$. Since
the different $h_{i}$ correspond to linear dependent vector columns of $A$
we find that the chirotope $\chi (i_{1},i_{2},i_{3})$ vanishes for the
different $h_{i}$. Hence, using (15) we find that the only nonvanishing
terms of $\chi (i_{1},i_{2},i_{3})$ are

\begin{equation}
\begin{array}{ccccccc}
124+ & 125+ & 126- & 127+ & 134- & 135+ & 136- \\ 
137- & 145+ & 146- & 157- & 167+ & 234- & 235+ \\ 
236+ & 237+ & 245+ & 247+ & 256+ & 267- & 346- \\ 
347- & 356+ & 357+ & 456+ & 457+ & 467- & 567+.
\end{array}
\label{17}
\end{equation}
On the other hand, for the different non-zero values of the octonions
structure constants $\psi _{i_{1}i_{2}i_{3}}$, we can choose any consistent
sign values for the different $h_{i}$. We shall make the following choice

\begin{equation}
\begin{array}{ccccccc}
123+ & 147- & 156- & 246+ & 257- & 345- & 367-.
\end{array}
\label{18}
\end{equation}
We observe that $\chi $ takes sign values different from zero precisely when
the corresponding $\psi $ takes sign values equal to zero and vice versa.
Therefore, this is another way of showing that $\chi $ and $\psi $ are
complementary concepts.

Now if we put together the sign set combinations (17) and (18) according to
(16) we get the sign values for $C_{i_{1}i_{2}i_{3}}\in \{-1,1\}$ which can
be identified with the chirotope of the oriented uniform matroid $\mathcal{M}%
(U_{3,7}).$ From (17) we can check by a direct computation that $F_{7}$ is
not orientable , while from (17) and (18) we can verify the orientability of 
$U_{3,7}$.

How is the above matroid-octonions scenario related to $G_{2}$-holonomy of $%
D=11$ supergravity? As it is known recent investigations of M-theory have
suggested to take $G_{2}$-holonomy seriously for the $7D$ manifold of $D=11$
supergravity. In particular, by a generalization of the self-duality
condition in four dimensions to eight dimensions Nishino and Rajpoot$^{23}$
have shown a mechanism that implements the $G_{2}-$symmetry in
supersymmetric models in $D\leq 8$. The interesting thing about this
mechanism is that it opens the possibility of obtaining the supersymmetric
vector multiplet with the full $SO(7)$ Lorentz covariance reduced to $G_{2}$
by dimensional reduction. The generalized self-duality condition used by
Nishino and Rajpoot can be derived from the definitions

\begin{equation}
F_{[mnr8]}\equiv \psi _{mnr}  \label{19}
\end{equation}
and 
\begin{equation}
F_{ijkl}\equiv \varphi _{ijkl},  \label{20}
\end{equation}
and by using (13). In fact, from (13), (19) and (20) it is found that $%
F_{\mu \nu \alpha \beta }$, with the Greek indices running from $1$ to $8$,
satisfies the self-duality formula

\begin{equation}
F^{\mu \nu \alpha \beta }=(1/4!)\epsilon ^{\mu \nu \alpha \beta \lambda
\gamma \rho \sigma }F_{\lambda \gamma \rho \sigma }.  \label{21}
\end{equation}
This extension of the octonionic relation (13) from $7D$ to $8D$ must have
his corresponding extension in matroid theory in connection with $F_{7}$. In
particular, this observation suggested the extension of the formula (16) to
the form

\begin{equation}
F_{\mu _{1}\mu _{2}\mu _{3}\mu _{4}}+\chi (\mu _{1},\mu _{2},\mu _{3},\mu
_{4})=C_{\mu _{1}\mu _{2}\mu _{3}\mu _{4}}.  \label{22}
\end{equation}
Our task is now to find now the mathematical meaning of $\chi (\mu _{1},\mu
_{2},\mu _{3},\mu _{4})$ and $C_{\mu _{1}\mu _{2}\mu _{3}\mu _{4}}.$ First
of all, since the chirotopes are defined in terms of the bases of a matroid,
it follows that $\chi (\mu _{1},\mu _{2},\mu _{3},\mu _{4})$ should
correspond to a four-rank matroid. Further, $\chi (\mu _{1},\mu _{2},\mu
_{3},\mu _{4})$ must have the property to be reduced to the chirotope $\chi
(i_{1},i_{2},i_{3})$ associated to $F_{7}.$ This means that such a four-rank
matroid should provide an extension of $F_{7}.$ In other words $F_{7}$
should be a minor of such a four-rank matroid. Moreover, the self-duality of 
$F_{\mu _{1}\mu _{2}\mu _{3}\mu _{4}}$ should imply a self-duality not only
for $\chi (\mu _{1},\mu _{2},\mu _{3},\mu _{4})$ but also for $C_{\mu
_{1}\mu _{2}\mu _{3}\mu _{4}}$. This means that the four-rank matroid must
be a self-dual matroid. Happily, there is a four rank matroid satisfying all
these requirements. In fact, the affine matroid $AG(3,2)$ is a self-dual
four-rank matroid and among its minors it has the matroid $F_{7}.$
Specifically, $AG(3,2)$ is not graphic, nor cographic, and every
single-element deletion is isomorphic to $F_{7}^{\ast }$ and every
single-element contraction is isomorphic to $F_{7}$. Furthermore, one finds
that the chirotope $\chi (\mu _{1},\mu _{2},\mu _{3},\mu _{4})$ associated
to $AG(3,2)$ is self dual and has all the desired properties.

In connection with $C_{\mu _{1}\mu _{2}\mu _{3}\mu _{4}},$ we already
mentioned that this object should be self-dual. It must also correspond to
the chirotope of a four rank matroid. Since $F_{\mu _{1}\mu _{2}\mu _{3}\mu
_{4}}\neq 0$ when $\chi (\mu _{1},\mu _{2},\mu _{3},\mu _{4})=0$ and $F_{\mu
_{1}\mu _{2}\mu _{3}\mu _{4}}=0$ when $\chi (\mu _{1},\mu _{2},\mu _{3},\mu
_{4})\neq 0$ we find that $C_{\mu _{1}\mu _{2}\mu _{3}\mu _{4}}$ should be
associated to a four-rank matroid with an underlying set $E=\{1,2,...,8\}$
and with its bases given by all of the four element subsets of $E$. The
candidate is, of course, the uniform matroid $U_{4,8}$ which is orientable$.$
In general, the uniform matroid $U_{p,q}$ has the self dual property $%
U_{p-q,q}=U_{p,q}$ and therefore we discover that $U_{4,8}$ is also
self-dual as required.

\bigskip

\smallskip\ 

\noindent \textbf{IV. GRAVITOID THEORY}

\bigskip

A physical theory that combines the concepts of matroid and graviton we
shall call it gravitoid theory. From this definition it turns out that the
connection of the Fano matroid $F_{7}$ and the affine matroid $AG(3,2)$ with 
$D=11$ supergravity discussed in the previous section can be considered as
part of the gravitoid theory.

Perhaps the simplest motivation to be interested in gravitoid theory may
follow from the heuristic observation that since the graviton can be
represented classically as a symmetric second-rank matrix tensor and a
matroid is a more general concept than a matrix one should expect that a
graviton-matroid connection may lead to an extension of the concept of the
graviton. (It seems to be reasonable to call such a generalized
gravitational structure the gravitoid.) The real motivation, however, may
arise from the idea of establishing a duality principle in M-theory. As we
mentioned in the introduction, duality is one of the most attractive
features in matroid theory. In a certain sense, matroid theory is the theory
of duality. Therefore, matroid theory offers an excellent mathematical
framework to incorporate a duality principle in M-theory. It turns out that
although the gravitoid theory may be interesting by itself, it can also
teach us, among other things, how to achieve such a quest.

Before we discuss the formal expected approach for the gravitoid theory, it
is first convenient to briefly mention the recently proposed mathematical
formalism of matroid bundle. This mathematical structure is the
combinatorial analogue of real vector bundles in which a ''base space'' is a
simplicial complex and the ''fibers'' are oriented matroids. The idea of
matroid bundles arose from the concept of combinatorial differential
manifolds introduced by MacPherson$^{14}$ in 1993 and evolved very rapidly
on several fronts. In particular, a close connection between matroid bundle
theory and characteristic classes has emerged.$^{24,25}$

Two main ideas make the matroid bundle structure possible: oriented matroids
and simplicial complexes. In order to define formally matroid bundles we
closely follow Anderson's work (see Ref. 15). Let $\xi :\mathcal{E}%
\rightarrow B$ be a real vector bundle, where $B$ is a $n-$dimensional base
space. We assume $B$ to be a triangulable compact space, i.e., $B$ has an
associated simplicial complex. For each point $b$ of $B$ the vectors $%
\{e_{1}(b),...,e_{n}(b)\}$ span the space $\xi ^{-1}(b)$. We can associate
an oriented matroid with underlaying set $E=\{1,2,...,n\}$ to the vectors $%
\{e_{1}(b),...,e_{n}(b)\}$.

A rank-$k$ matroid bundle is a pair $(S_{B},\mathcal{M})$ where $S_{B}$ is a
partially ordered set and $\mathcal{M}$ is a rank-$k$ oriented matroid
associated to each element $b$ so that $\mathcal{M}(b)$ weakly maps to $%
\mathcal{M}(b^{\prime })$ whenever $b\geq b^{^{\prime }}$. Intuitively, and
in analogy to the case of a real vector bundle, $S_{B}$ corresponds to the
base space $B$ and the oriented matroid $\mathcal{M}$ to the fiber of the
bundle $\mathcal{E}$ (For a complete exposition see Ref. 15).

In particular, in the case of tangent matroid bundles one associates an
oriented matroid to each tangent space and one realizes a differentiable
manifold by a combinatorial differential manifold. In what follows, we shall
focus our attention on this special kind of matroid bundle.

In order to introduce the gravitoid concept it is first convenient to
recall, in brief, how the metric is introduced in a tangent bundle. Consider
a tangent bundle over the differentiable manifold $B$:

\begin{equation}
T(B)=\bigcup_{b\in B}T_{b}(B).  \label{23}
\end{equation}
Here, the tangent space $T_{b}(B)$ plays the role of the fiber of $T(B)$.
(We assume that $B$ is a triangulable manifold.) Of course, $T_{b}(B)$ is a
vector space and consequently $T(B)$ provides an example of a real vector
bundle. The dual of $T(B)$ is the cotangent bundle $T^{\ast }(B)$ defined in
the form

\begin{equation}
T^{\ast }(B)=\bigcup_{b\in B}T_{b}^{\ast }(B),  \label{24}
\end{equation}
where $T_{b}^{\ast }(B)$ is the dual vector space of $T_{b}(B)$. Thus,
whereas $T_{b}(B)$ is a vector space of contravariant vectors, $T_{b}^{\ast
}(B)$ is a vector space of covariant vectors (1-forms). Moreover, it can be
proved that the dimension of $T_{b}^{\ast }(B)$ is equal to the dimension of 
$T_{b}(B)$.

Take $\frac{\partial }{\partial b^{\mu }}$ $(1\leq \mu \leq n)$ as the base
vectors of $T_{b}(B)$ and $db^{\mu }$ as the corresponding bases (one forms)
of $T_{b}^{\ast }(B)$. Consider a vector $V^{\mu }\frac{\partial }{\partial
b^{\mu }}$ and its dual one-form $V_{\mu }db^{\mu }$. The tangent space
isomorphism

\begin{equation}
g:T_{b}(B)\rightarrow T_{b}^{\ast }(B)  \label{25}
\end{equation}
has the component representation

\begin{equation}
V_{\mu }=g_{\mu \nu }V^{\nu }.  \label{26}
\end{equation}
Expressions (25) or (26) can be used to define an inner product $\langle
\omega ,V\rangle :T_{b}^{\ast }(B)\otimes T_{b}(B)\rightarrow R$, where $R$
is the set of real numbers. In component representation this leads to

\begin{equation}
\langle \omega ,V\rangle =\omega _{\mu }V^{\mu },  \label{27}
\end{equation}
which by virtue of (26) can also be written as

\begin{equation}
\langle \omega ,V\rangle =g_{\mu \nu }\omega ^{\mu }V^{\nu }.  \label{28}
\end{equation}
Usually the matrix $g_{\mu \nu }$ is called a metric if it is symmetric and
positive definite, so $\langle V,V\rangle $ has the meaning of square norm
of $V$. However, in the case of a pseudo Riemannian space the positive
definite condition can be dropped and we can still call the matrix $g_{\mu
\nu }$ a metric (or pseudometric). Moreover, in nonsymmetric gravitational
theories the matrix $g_{\mu \nu }$ does not even satisfy the symmetric
condition. Therefore, in order to be as general as possible we shall call a
metric just the isomorphism (25), without specifying any additional
condition.

Let us now return to our goal of finding a mathematical structure for the
gravitoid system. In the case of a tangent matroid bundle, $T_{b}(B)$ is
replaced by an oriented matroid $\mathcal{M}_{T_{b}}$ and the base space is
replaced by a simplicial complex $S_{B}$. Just like one can associate to $%
T_{b}(B)$ the dual tangent space $T_{b}^{\ast }(B),$ one can associate to $%
\mathcal{M}_{T_{b}}$ the dual oriented matroid $\mathcal{M}_{T_{b}}^{\ast }.$
Thus, one should expect that the isomorphism $g:T_{b}(B)\rightarrow
T_{b}^{\ast }(B)$ leads to the corresponding isomorphism

\begin{equation}
\mathcal{G}_{g}:\mathcal{M}_{T_{b}}\rightarrow \mathcal{M}_{T_{b}}^{\ast }.
\label{29}
\end{equation}
In general if an oriented matroid is realizable, its dual is also
realizable. Therefore, if $\mathcal{M}_{T_{b}}$ is realizable then $\mathcal{%
M}_{T_{b}}^{\ast }$ is also realizable. Let us denote by $A$ the $k\times n$
matrix over a field $R$ representing the vector matroid $\mathcal{M}%
_{T_{b}}[A].$ The ground set $E$ of $\mathcal{M}_{T_{b}}[A]$ are the columns
of $A$. Similarly let us denote by $A^{\ast }$ the corresponding matrix
associated to the dual vector matroid $\mathcal{M}_{T_{b}}^{\ast }[A].$ In
this context the map (29) has the component representation

\begin{equation}
a_{i}=\mathcal{G}_{ij}a^{j},  \label{30}
\end{equation}
where $a^{j}\in A$ and $a_{i}\in A^{\ast }$.

Now, the question arises whether there is a relation between the two maps $g$
and $\mathcal{G}$. The map $\mathcal{G}$ can be understood as the
combinatorial analogue of the metric $g$. For this reason we shall call $%
\mathcal{G}_{ij}$ the matroid metric. Since the simplicial complex $S_{B}$
is homeomorphic to the smooth manifold $B,$ i.e., $\lambda :S_{B}\rightarrow
B$ is a homeomorphism, we have the induced map $\eta :\mathcal{M}%
_{T_{b}}\rightarrow V_{b}(B)$, where $V_{b}(B)$ is a finite set of vectors
in $T_{b}(B)$. Hence, (25) and (29) imply that there must exist a map $\eta
^{\ast }:\mathcal{M}_{T_{b}}^{\ast }\rightarrow V_{b}^{\ast }(B)$, where $%
V_{b}^{\ast }(B)$ is a finite set of covectors in $T_{b}^{\ast }(B)$ and $%
\eta ^{\ast }=g\eta \mathcal{G}^{-1}$. This result establishes the
connection between $g$ and $\mathcal{G}$.

Our next step is to associate a gravitational theory to the matroid metric $%
\mathcal{G}_{ij}$. For this purpose we first need to introduce the
combinatorial analogue of a connection and curvature. At this respect we
shall relay on Ref. 24. The idea is to define 1-cocycle $\Theta $ and
2-cocycle $\rho ^{\ast }\Omega =\delta \Theta $ on each local system on $%
S_{B}$. It is worth mentioning that, in order to find a combinatorial
formula for the Pontrjagin classes, Gelfand and MacPerson consider $\Theta $
and $\Omega $ as the combinatorial analogue of connection and curvature of $%
B $, respectively.

The quest now is to find a relation between $\mathcal{G}$ and $\Omega $. In
principle, we can achieve this by applying a Palatini method to certain
proposed action $I=I(\mathcal{G},\Omega ).$ The question is, therefore, what
is the form of $I(\mathcal{G},\Omega )$? A partial answer to this question
may come from the link between matroid theory and Chern-Simons theory found
in Ref. 7. For this reason let us briefly review the main ideas of Ref. 7.

It is known that if we choose $B^{3}=S^{3},$ $G=SU(2)$ and $\rho _{r}=C^{2}$
for all the link components then the Witten's partition function

\begin{equation}
Z(L,k)=\int DA\exp (S_{cs})\prod\limits_{r=1}^{n}W(L_{r},\rho _{r}),
\label{31}
\end{equation}
where $S_{CS}$ is the Chern-Simons action

\begin{equation}
S_{CS}=\frac{k}{2\pi }\int_{M^{3}}Tr(A\wedge dA+\frac{2}{3}A\wedge A\wedge A)
\label{32}
\end{equation}
and $W(C_{i},\rho _{i})$ is the Wilson line

\begin{equation}
W(L_{r},\rho _{r})=Tr_{\rho _{r}}P\exp (\smallint
_{L_{r}}A_{i}^{a}T_{a}dx^{i})  \label{33}
\end{equation}
reproduces the Jones polynomial

\begin{equation}
Z(L,k)=V_{L}(t).  \label{34}
\end{equation}
Here, $A=A_{i}^{a}T_{a}dx^{i}$, with $T_{a}$ the generators of the Lie
algebra of a gauge group $G$ and the symbol $P$ means the path-ordering
along the knots $L_{r}.$ The parameter $t$ is 
\begin{equation}
t=e^{\frac{2\pi i}{k}}  \label{35}
\end{equation}
and $V_{L}(t)$ denotes the Jones polynomial satisfying the skein relation;

\begin{equation}
t^{-1}V_{L_{+}}-tV_{L_{-}}=(\sqrt{t}-\frac{1}{\sqrt{t}})V_{L_{0}},
\label{36}
\end{equation}
where $L_{+},L_{-}$ and $L_{0}$ are the standard notation for overcrossing,
undercrossing and zero crossing.

On the other hand, Thistlethwaite$^{26}$ showed that if $L$ is an
alternating link and $\mathit{G}(L)$ the corresponding planar graph, then
the Jones polynomial $V_{L}(t)$ is equal to the Tutte polynomial $T_{\mathit{%
G}}(-t,-t^{-1})$ up to a sign and a factor power of $t.$ Specifically, we
have

\begin{equation}
V_{L}(t)=(-t^{\frac{3}{4}})^{w(L)}t^{\frac{-(r-n)}{4}}T_{\mathit{G}%
}(-t,-t^{-1}),  \label{37}
\end{equation}
where $w(L)$ is the writhe and $r$ and $n$ are the rank and the nullity of $%
\mathit{G,}$ respectively$.$ Here, $V_{L}(t)$ is the Jones polynomial of the
alternating link $L$. The Tutte polynomial associated to each graph $\mathit{%
G}$ is a polynomial $T_{\mathit{G}}(x,x^{-1})$ with the property that if $%
\mathit{G}$ is composed solely of isthmus and loops then $T_{\mathit{G}%
}(x,x^{-1})=x^{I}x^{-l},$ where $I$ is the number of isthmuses and $l$ is
the number of loops. The polynomial $T_{\mathit{G}}$ satisfies the skein
relation

\begin{equation}
T_{\mathit{G}}=T_{\mathit{G}^{^{\prime }}}+T_{\mathit{G}^{^{\prime \prime
}}},  \label{38}
\end{equation}
where $\mathit{G}^{\prime }$ and $\mathit{G}^{\prime \prime }$ are obtained
by deleting and contracting respectively an edge that is neither a loop nor
an isthmus of $\mathit{G}$.

A theorem due to Tutte allows to compute $T_{\mathit{G}}(-t,-t^{-1})$ from
the maximal trees of $\mathit{G}$. In fact, Tutte proved that if $\mathcal{B}
$ denotes the set of maximal trees in a graph $\mathit{G}$, $i(\mathcal{B})$
denotes the number of internally active edges in $\mathit{G,}$ and $e(B)$
refers to the number of the externally active edges in $\mathit{G}$ (with
respect to a given maximal tree $B\in \mathcal{B}$) then the Tutte
polynomial is given by the formula

\begin{equation}
T_{\mathit{G}}(-t,-t^{-1})=\sum_{B\subseteq \mathcal{B}}x^{i(B)}x^{-e(B)},
\label{39}
\end{equation}
where the sum is over all elements of $\mathcal{B}$. It is worth mentioning
that the Tutte polynomial $T_{\mathit{G}}(-t,-t^{-1})$ computed according to
(39) uses the concept of a graphic matroid $M(\mathit{G)}$ defined as the
pair $(E,\mathcal{B})$, where $E$ is the set of edges of $\mathit{G}$.

Therefore, we have the connections $M(\mathit{G)\leftrightarrow }T_{\mathit{G%
}}(-t,-t^{-1})\leftrightarrow V_{L}(t)\leftrightarrow Z(L,k)$. These bridges
allow to transfer information from $M(\mathit{G)}$ to $Z(L,k)$ and conversely%
$.$ In particular, if we change the notation $T_{\mathit{G}%
}(-t,-t^{-1})\rightarrow T_{M(\mathit{G)}}(t)$ and $Z(L,k)\rightarrow Z_{M(%
\mathit{G)}}(k)$ one discovers that the duality property of the Tutte
polynomial

\begin{equation}
T_{\mathit{G}}(-t,-t^{-1})=T_{\mathit{G}^{\ast }}(-t^{-1},-t)  \label{40}
\end{equation}
can be expressed as

\begin{equation}
T_{M(\mathit{G)}}(t)=T_{M^{\ast }(\mathit{G)}}(t^{-1})  \label{41}
\end{equation}
and consequently from (34) and (37) we find that the partition function $%
Z_{M(\mathit{G)}}(k)$ gets the dual property

\begin{equation}
Z_{M(\mathit{G)}}(k)=Z_{M^{\ast }(\mathit{G)}}(-k),  \label{42}
\end{equation}
(see Ref. 7 for details). This expression is of special relevance because it
is exactly the kind of symmetry that one may expect in M-theory. Writing $%
k=\ln \tau $, where $\tau $ is a constant parameter (perhaps complex) one
finds that the symmetry (42) in $Z_{M(\mathit{G)}}(k)$ is related to the
S-duality symmetry $\tau \rightarrow -\frac{1}{\tau }$ which is one of the
key symmetries linking different superstring theories.

In the gravitoid theory one should expect similar, but not quite the same
construction as the matroid-Chern-Simons approach. First of all, the
Chern-Simons action in (31) is not written in abstract combinatorial terms.
(Perhaps, here again the Gelfand-MacPerson$^{24}$ procedure could be
useful.) Secondly, the matroid theory enters in the scenario via graph
theory. But graph theory is just an example of the simplicial complex
formalism, so one needs the analogue for the Tutte polynomial for higher
dimensional simplexes and, as far as we know, no such a mathematical
structure is available in the mathematical literature. Nevertheless, let us
just simply write the expected abstract symmetry that the gravitoid
partition function should have, namely

\begin{equation}
Z_{\mathcal{M}_{T_{b}}}(\mathcal{G},\Omega )=Z_{\mathcal{M}_{T_{b}}^{\ast }}(%
\mathcal{G}^{\ast },\Omega ^{\ast }),  \label{43}
\end{equation}
where $\Omega ^{\ast }$ is the 2-cocycle associated to $\mathcal{G}^{\ast }$%
. Since every matroid $\mathcal{M}$ satisfies

\begin{equation}
\mathcal{M}^{\ast \ast }=\mathcal{M}  \label{44}
\end{equation}
we should have $\mathcal{G}^{\ast }=\mathcal{G}^{-1}$. The fact that any
matroid has its corresponding dual matroid is translated as a dual symmetry
of the partition function $Z_{\mathcal{M}_{T_{b}}}(\mathcal{G},\Omega )$.
This is a remarkable consequence of using matroid theory as a mathematical
framework. In order to have a complete picture in the expression (43) it is
necessary to find the combinatorial analogue of the concept of classical
action. Although this seems to be not a trivial problem, we can speculate
that the answer should be something similar to the combinatorial formula for
the Pontrjagin classes due to Gelfand and MacPherson.$^{24}$ Oriented
matroids and the simplicial complex formalism allowed these authors to find
the combinatorial analogue of the Chern-Weil formula

\begin{equation}
\tilde{p}_{i}(E)=(-1)^{i}\pi _{\star }\Omega ^{(e-2+2i)}.  \label{45}
\end{equation}
Here, $\pi _{\star }$ represents an integration over a fiber. In Chern-Weil
theory the curvature is defined by $\rho ^{\star }\Omega =d\Theta ,$ where $%
\Theta $ is a one form induced by the connection on the bundle $E$. In (45) $%
e$ is the fiber dimension of $E$. The combinatorial analogue of (45)
proposed by Gelfand and MacPherson is

\begin{equation}
\tilde{p}_{i}(X)\smallfrown \lbrack X]=(-1)^{i}\pi _{\star }(\frac{1}{2}%
\Omega )^{(n-2+2i)}\smallfrown \phi ),  \label{46}
\end{equation}
where $\phi $ is a fixing cycle for the simplicial complex $X$ (For details
and further information we refer the reader to the original Ref. 24 of
Gelfand and MacPherson.)

Since in matroid theory we have that any two connected matroids $\mathcal{M}%
_{1}$ and $\mathcal{M}_{2}$ satisfy the expression $(\mathcal{M}_{1}\oplus 
\mathcal{M}_{2})^{\ast }=\mathcal{M}_{1}^{\ast }\oplus \mathcal{M}_{2}^{\ast
}$ we discover the remarkable result that

\begin{equation}
Z_{\mathcal{M}_{1T_{b}}}=Z_{\mathcal{M}_{1T_{b}}^{\ast }}  \label{47}
\end{equation}
and

\begin{equation}
Z_{\mathcal{M}_{2T_{b}}}=Z_{\mathcal{M}_{2T_{b}}^{\ast }}  \label{48}
\end{equation}
if and only if

\begin{equation}
Z_{\mathcal{M}_{1T_{b}}\oplus \mathcal{M}_{2T_{b}}}=Z_{(\mathcal{M}%
_{1T_{b}}\oplus \mathcal{M}_{2T_{b}})^{\ast }}.  \label{49}
\end{equation}

Besides its importance in gravitoid theory, the equation (43) may motivate a
similar construction for M-theory. As it was mentioned before the five known
superstring theories in 9+1 dimensions can be understood via duality as a
different vacuum of the M-theory. Therefore, one should expect that the
partition function associated to M-theory should be invariant under certain
duality symmetry such as (42) or (43).

\bigskip

\smallskip\ 

\noindent \textbf{VI. COMMENTS}

\bigskip

Matroid bundle is a remarkable mathematical structure that promises many
interesting applications not only in mathematics but also in physics. In the
present work we have shown the possibility of using matroid bundle as a
framework for a new gravitational theory called gravitoid theory. The
essential idea in this theory is to combine the concepts of graviton and
matroid. One of the hopes is that gravitoid theory shows us how to develop
M-theory from M(atroid)-theory. In essence we have proposed a partition
function with a duality symmetry for the gravitoid system which may suggest
a similar structure for M-theory.

Our procedure was focused on realizable tangent matroid bundles. However,
since not all of the oriented matroids are realizable there must be more
general tangent matroid bundles and therefore a more general gravitoid
theory. Mathematically, a similar question has been considered in Ref. 14
where it was raised the question whether all combinatorial differential
manifolds are topological manifolds.

One of the attractive features of matroid bundle is that it offers the
possibility of using new combinatorial techniques to study characteristic
classes. In particular, it has been proved that matroid bundles may be the
basis for new studies on characteristic classes of differential manifolds.
Examples of this fact is the Gelfand-MacPherson's combinatorial formula$%
^{24} $ for the Pontrjagin classes and the Anderson-Davis's
combinatorialization$^{25}$ of the Stiefel-Whitney classes and Euler
classes. These characteristic classes play an essential role in theoretical
physics. For instance, Stiefel-Whitney classes are useful to determine the
spin structure of a smooth manifold and Euler classes and Pontrjagin classes
are useful topological invariants in the formulation of the
MacDowell-Mansouri's formalism (see Refs. 27 and references there in) in
supergravity. These observations suggest and motivate the search of a link
between the gravitoid theory and these combinatorial characteristic classes.

Gravitoid theory can be understood as the combinatorialization of the
bosonic sector of supergravity in any dimension and in particular in $D=11$.
In view of the MacPherson%
\'{}%
s conjecture$^{14,16}$ that all characteristic classes of a vector bundle
should have their corresponding characteristic classes in matroid bundle one
should expect combinatorialization of Chern-Simons theory and topological
gravity$^{28}$. From this perspective it is tempting to speculate about the
possibility that gravitoid theory may contain, in fact, the
combinatorialization of topological gravity.

There are many other open questions. Is gravitoid theory an alternative for
quantum gravity? What is the supersymmetrization of the gravitoid theory?
What is the precise relation between the gravitoid theory and $D=11$
supergravity? It seems that the combinatorialization of the Pontrjagin
classes implies an exotic combinatorial $7D$ sphere. Is the gravitoid theory
related to this exotic structure via $D=11$ supergravity? Our hope is that
the present work stimulates a research to answer these and other related
questions.

\bigskip

\noindent \textbf{Acknowledgment:} We would like to thank C. Austin for
calling our attention about the Anderson and Davis's references. We also
like to thank to J. Saucedo and G. Compoy for helpful comments.


\bigskip

\begin{thebibliography}{99}
\bibitem{1}  P. K. Townsend, ``Four lectures on M-theory,'' \textit{%
Proceedings of the ICTP on the Summer School on High Energy Physics \ and
Cosmology}, June 1996, hep-th/9612121.

\bibitem{2}  M. J. Duff, Int. J. Mod. Phys. A\textbf{\ 11,} 5623 (1996),
hep-th/9608117.

\bibitem{3}  P. Horava and E. Witten, Nucl. Phys. B\textbf{\ 460,} 506
(1996).

\bibitem{4}  M. Green, V. Schwarz and E. Witten, \textit{Superstrings Theory}
(Cambridge University Press, Cambridge, 1987) Vol I and II; M. Kaku, \textit{%
Introduction to Superstrings} (Spring-Verlag, Berlin, 1990).

\bibitem{5}  H. Whitney, Am. J. Math. \textbf{57,} 509\ (1935).

\bibitem{6}  J. A. Nieto, Rev. Mex. Fis. \textbf{44,} 358 (1998).

\bibitem{7}  J. A. Nieto and M. C. Marin, J. Math. Phys. \textbf{41,} 7997
(2000).

\bibitem{8}  J. A. Nieto, ``Searching for a connection between matroid
theory and string theory,'' hep-th/0212100.

\bibitem{9}  J. G. Oxley, \textit{Martroid Theory}, (Oxford University
Press, New York, 1992)

\bibitem{10}  D. J. A. Welsh, \textit{Martroid Theory}, (Academic, London,
1976).

\bibitem{11}  R. J. Wilson, \textit{Introduction to Graph Theory}, 3rd ed.
(Wiley, New York, 1895).

\bibitem{12}  J. P. S. Kung, \textit{A Source Book in Matroid Theory},
(Birkhauser, Boston, 1986).

\bibitem{13}  K. Ribnikov, \textit{An\'{a}lisis Combinatorio}, (Editorial
Mir, Mosc\'{u}, 1988)

\bibitem{14}  R. D. Macpherson, ``Combinatorial differential manifolds: a
symposium in honor of John Milnor's sixtieth birthay,'' pp. 203-221 in
Topological methods on modern mathematics (Stony Brook, NY, 1991), edited by
L. H. Goldberg and A. Phillips, Houston, 1993.

\bibitem{15}  L. Anderson, New Perspectives. in Geom. Comb. \textbf{38}, 1
(1999).

\bibitem{16}  D. Biss, ``Some applications of oriented matroids to
topology,'' PhD. thesis, MIT, 2002.

\bibitem{17}  Babson, ``A combinatorial flag space,'' PhD. thesis, MIT,
(1993).

\bibitem{18}  A. Bjorner, M. Las Verganas, N. White and G. M. Ziegler, 
\textit{Oriented Martroids}, (Cambridge University Press, Cambridge, 1993).

\bibitem{19}  H. Whitney, Trans. Amer. J. Math. \textbf{34}, 339 (1932)

\bibitem{20}  L. Gutierrez-Novoa, Pacific J. Math. \textbf{15}, 1337 (1965).

\bibitem{21}  F. Englert, Phys. Lett. B\textbf{\ 119,} 339 (1982); F. Gursey
and C. Tze, Phys. Lett. B \textbf{127}, 191 (1983).

\bibitem{22}  I. L. Kantor and A.S. Solodovnikov, \textit{Hypercomplex
Numbers; An Elementary Introduction to Algebras,} (Spring Verlag, New York,
1989); J. C. Baez, Bull. Amer. Math. Soc. \textbf{39}, 145 (2002).

\bibitem{23}  H. Nishino and S. Rajpoot, ``Octonions, G(2) symmetry,
generalized selfduality and supersymmetries in dimensions D less than or
equal 8,'' hep-th/0210132.

\bibitem{24}  I. M. Gelfand and R. D. Macpherson, Bull. Amer. Math. Soc. 
\textbf{26}, 304 (1992).

\bibitem{25}  L. Anderson and J. F. Davis, ``Mod 2 Cohomolgy of
Combinatorial Grassmannians,'' math.GT/9911158.

\bibitem{26}  M. Thistlethwaite, Topology \textbf{26}, 297 (1987).

\bibitem{27}  H. Garcia-Compean, J. A. Nieto, O. Obregon and C. Ramirez,
Phys. Rev. D \textbf{59}, 124003 (1999); J. A. Nieto and J. Socorro Phys.
Rev. D \textbf{59}, 041501 (1999); J. A. Nieto, J. Socorro and O. Obregon,
Phys. Rev. Lett.\textbf{76}, 3482 (1996).

\bibitem{28}  E. Witten, Phys. Lett. B \textbf{206}, 601 (1988).
\end{thebibliography}
\end{document}